# Hole Detection for Increasing Coverage in Wireless Sensor Network Using Triangular Structure

Shahram Babaie[1] and Seyed Sajad Pirahesh[2]

[1] Department of Computer engineering, Tabriz-Branch, Islamic Azad University, Tabriz, Iran

[2] Department of Computer engineering, Tabriz-Branch, Islamic Azad University, Tabriz, Iran

**Abstract**

The emerging technology of wireless sensor network (WSN) is expected to provide a broad range of applications, such as battlefield surveillance, environmental monitoring, smart spaces and so on. The coverage problem is a fundamental issue in WSN, which mainly concerns with a fundamental question: How well a sensor field is observed by the deployed sensors? Mobility is exploited to improve area coverage in a kind of hybrid sensor networks. The main objective for using mobile sensor nodes is to heal coverage holes after the initial network deployment, when designing a hole healing algorithm, the following issues need to be addressed. First, how to decide the existence of a coverage hole and how to estimate the size of a hole. Second, what are the best target locations to relocate mobile nodes to repair coverage holes? We use the triangular oriented diagram (HSTT) for aim to goal where its simple ,have low calculation among construction and it is great to calculate the size of hole exactly .

## 1. Introduction

Recent advances in micro-electro-mechanical systems, embedded processors, and wireless communications have led to the emergence of Wireless sensor networks (WSNs), which consist of a large number of sensing devices each capable of sensing, processing and transmitting environmental information. Applications of WSNs include battlefield surveillance, environmental monitoring, biological detection, smart spaces, industrial diagnostics, and so on [1].

A fundamental issue in WSNs is the coverage problem [2, 3]. The coverage problem is heavily dependent on the coverage model of individual sensor and the locations of the deployed sensor nodes. Sensor coverage model can be considered as a measure of the quality of service of sensor's sensing function and is subject to a wide range of interpretations due to a large variety of sensors and applications. In the literature, a widely used sensor coverage model is the sensing disk model where a sensor can cover a disk centered at itself with a radius equal to a fixed sensing range. Network sensing coverage on the other hand can be considered as a collective measure of the quality of service provided by sensor nodes at different geographical locations. In many cases, we may interpret the coverage concept as a non-negative mapping between the space points (of a sensor field) and the sensor nodes (of a deployed sensor network). For example, given the sensing disk model, the area (space points) covered by a set of sensors is the union of their sensing disks.

Wireless sensors can be either deterministic placed or randomly deployed in a sensor field. Deterministic sensor placement can be applied to a small to medium sensor network in a friend environment. When the network size is large or the sensor field is remote and hostile, random sensor deployment might be the only choice, e.g., scattered from an aircraft. It has been shown that a critical sensor density exists beyond which a sensor field can be completely covered almost surely in every random deployment [4, 5]. To guarantee complete coverage in one random deployment, it is often assumed that the number of scattered sensors is more than that required by the critical sensor density. However, this normally requires a great number of sensor nodes to be deployed another way to improve network coverage is to leverage mobile sensor nodes. Mobile sensor nodes are equipped with locomotive platforms and can move around after initial deployment, for example, the mobile sensor nodes Robomote [6] and iMouse [7]. Although in general a mobile sensor node is more expensive than its stationary compeer, it can serve much functionality such as a data relay or collector, and can greatly improve many network performances such as enhancing timeliness of data report. In this article, our focus is to healing coverage hole using genetic algorithm with minimize total movement of mobile sensor.

## 2. RELATED WORK

Bang Wang, Hock Beng Lim, Di Ma with article [12] proposed:

### 2.1 Hole detection and hole size estimation

Voronoi diagram can be used to detect a coverage hole and calculate the size of a coverage hole [8, 9]. A Voronoi diagram for N sensors $s_1, s_2,…,s_N$ in a plane is defined as the subdivision of the plane into N cells each for one sensor, such that the distance between any point in a cell and the sensor of the cell is closer than that distance between this point and any other sensors. Two Voronoi cells meet along a Voronoi edge and a sensor is a Voronoi neighbor of another sensor if they share a Voronoi edge. We refer the reader to [10] for more discussions on Voronoi diagram and its applications.

A Voronoi diagram is first constructed for all stationary sensor nodes, assuming that each node knows its own and its neighbors' coordinates. Wang et al. [9] proposes a localized construction algorithm to construct a local Voronoi diagram: Each node constructs its own Voronoi cell by only considering its 1-hop neighbors. After the local Voronoi diagram construction, the sensor field is divided into sub regions of Voronoi cells and each stationary node is within a Voronoi cell. A node is a Voronoi neighbor of another one if they share a Voronoi edge. Fig. 2 illustrates a Voronoi diagram in a bounded sensor field, where the boundaries of the sensor field also contribute to a Voronoi cell. According to the property of a Voronoi diagram, all the points within a Voronoi cell are closest to only one node that lies within this cell. Therefore, if some points of a Voronoi cell are not covered by its generating node, these points will not be covered by any other sensor and contribute to coverage holes. If a sensor covers all of its Voronoi cell's vertices, then there are no uncovered points within its Voronoi cell; otherwise, uncovered points exist within its Voronoi cell.

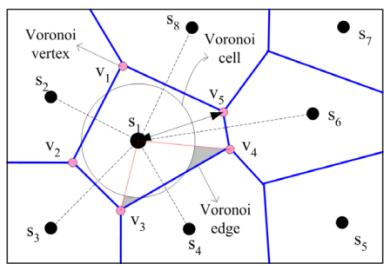

Fig. 1 Illustration of using Voronoi diagram to detect a coverage hole and decide the hole size.

### 2.2 Destination selection

After deciding the existence of a coverage hole and its size, a stationary node needs to decide the number of mobile nodes and the target locations of these mobile nodes to heal its holes. Ghosh [8] proposes that for each Voronoi vertex, one mobile node should be used to heal the coverage hole around this Voronoi vertex, if the size of its coverage hole within the Voronoi cell is larger than a threshold.

Wang et al. [13] convert the sensor movement problem into a maximum weight maximum-matching problem to decide which mobile node should move to which target location.

## 3. OUR WORK

We assume there is an environment like fig 1. That fill with stationary sensors at the first fore implement. In our diagram (HSTT) we connect the center of sensors sensing to gather with condition of made a triangular every three adjacent sensors.

Our diagram can be used to detect a coverage hole and calculate the size of a coverage hole

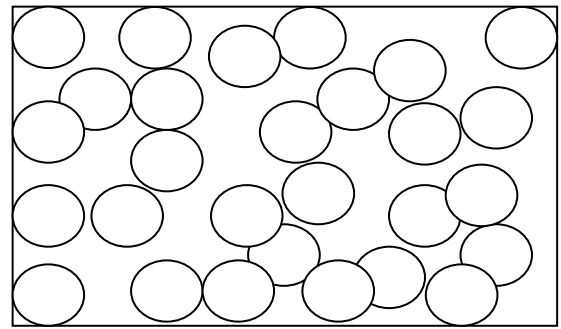

Fig.2 Initial deployment.

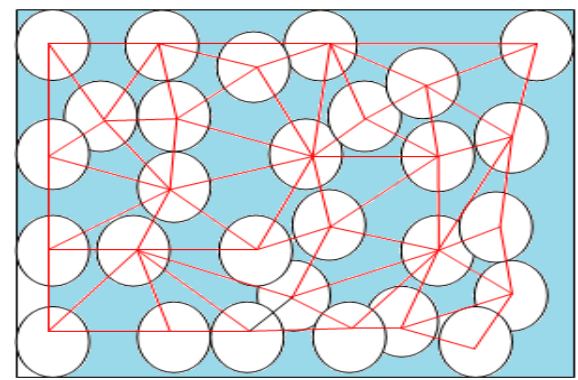

Fig.3 Constraction the triangular oreinted structure.

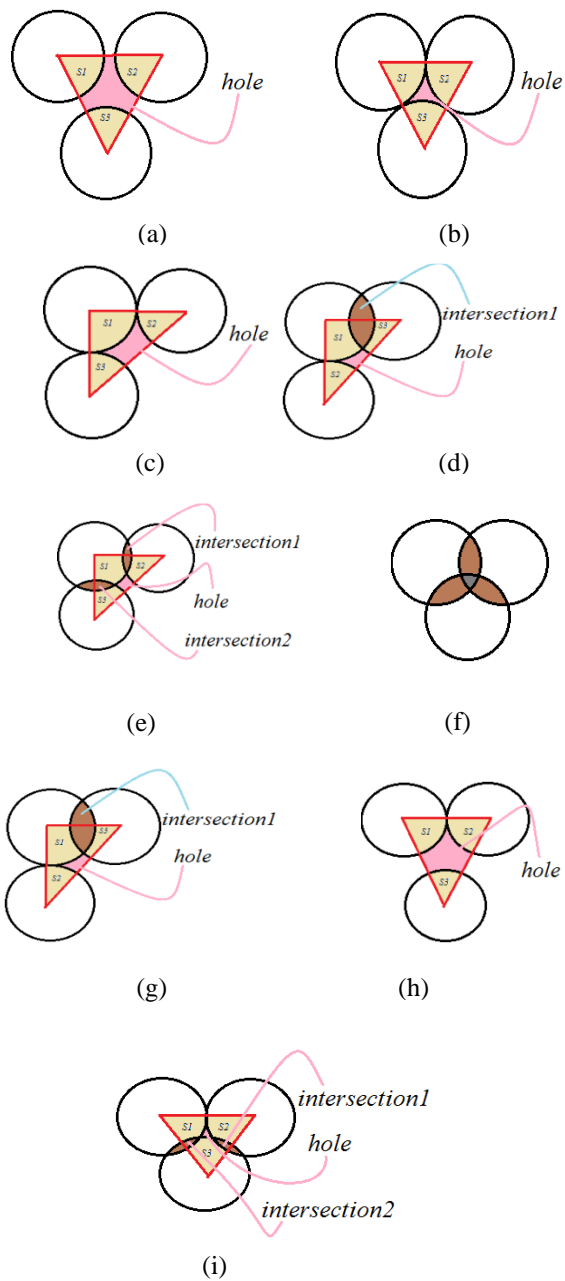

Fig.4 The total possible sates with three sensors.

Relation to area of the triangle.

Let A be the triangle's area and let a, b and c, be the lengths of its sides. By Heron's formula, the area of the triangle is

$$area = A$$
$$= \frac{1}{4}\sqrt{(a+b+c)(a-b+c)(b-c+a)(c-a+b)}$$

$$= \sqrt{s(s-a)(s-b)(s-c)} \qquad (1)$$

Where $s = \frac{a+b+c}{2}$ is the semi perimeter.

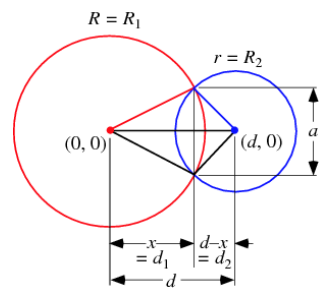

Fig.5 Intersection between two circles

Let two circles of radii **R** and **r** and centered at (0,0) and (d,0) intersect in a region shaped like an asymmetric lens. The equations of the two circles are

$$x^2 + y^2 = R^2 \qquad (2)$$
$$(x-d^2) + y^2 = r^2 \qquad (3)$$

Combining (1) and (2) gives

$$(x-d)^2 + (R^2 - x^2) = r^2 \qquad (4)$$

Multiplying through and rearranging gives

$$x^2 - 2dx + d^2 - x^2 = r^2 - R^2 \qquad (5)$$

Solving for **x** results in

$$x = \frac{d^2 - r^2 + R^2}{2d} \qquad (6)$$

The chord connecting the cusps of the lens therefore has half-length **y** given by plugging **x** back in to obtain

$$y^2 = R^2 - x^2 = R^2 - (\frac{d^2 - r^2 + R^2}{2d})^2$$
$$= \frac{4d^2 R^2 - (d^2 - r^2 + R^2)^2}{2d} \qquad (7)$$

Solving for **y** and plugging back in to give the entire chord length a=2 then gives

$$a = \frac{1}{d}\sqrt{4d^2R^2 - (d^2 - r^2 + R^2)^2} \quad (8)$$

$$= \frac{1}{2}\sqrt{(-d+r-R)(-d-r+R)(-d+r+R)(d+r+R)}$$

This same formulation applies directly to the sphere-sphere intersection problem To find the area of the asymmetric "lens" in which the circles intersect, simply use the formula for the circular segment of radius $R'$ and triangular height $d'$

$$A(R',d') = R'^2 \cos^{-1}(\frac{d'}{R'}) - d'\sqrt{R'^2 - d'^2} \quad (9)$$

twice, one for each half of the "lens." Noting that the heights of the two segment triangles are

$$d_1 = x = \frac{d^2 - r^2 - R^2}{2d} \quad (10)$$

$$d_2 = d - x = \frac{d^2 + r^2 - R^2}{2d}$$

The result is

$$A = A(R, d_1) + A(r, d_2) \quad (11)$$

$$= r^2 \cos^{-1}(\frac{d^2 + r^2 - R^2}{2dr}) - R^2 \cos^{-1}(\frac{d^2 - r^2 + R^2}{2dR}) \quad (12)$$

$$-\frac{1}{2}\sqrt{(d+r-R)(-d+r+R)(d-r+R)(d+r+R)}$$

Now we calculate the area fig 3.

**3.a**

If a, b, c>2R

$$s_h = s_\Delta - (s_1 + s_2 + s_3)$$

We know the total angular of triangular is:

$$\hat{\alpha} + \hat{\beta} + \hat{\zeta} = 1\hat{8}0$$

So

$$(s_1 + s_2 + s_3) = \pi R^2 \quad (13)$$

According to equation 1 we have

$$s_h = \sqrt{(a+b+c)(a-b+c)(b-c+a)(c-a+b)} - \pi R^2 \quad (14)$$

**3.b**

If a=b=c=2R

The area of hole calculates like section 3.a.

**3.c**

If a=b or a=c or b=c

The area of hole calculate like section 3.a .

**3.d**

$$s_h = s_\Delta - (s_1 + s_2 + s_3) + s_{\text{int}er\sec t}$$

According to equation 11 and By assume r=R we have

$$s_{\text{int}er\sec t} = R^2 \cos^{-1}(\frac{d^2}{2dR}) - R^2 \cos^{-1}(\frac{d^2}{2dR}) - \frac{1}{2}\sqrt{(-d+2R)d^2(d+2R)} \quad (15)$$

**3.e**

$$s_h = s_\Delta - (s_1 + s_2 + s_3) + \frac{1}{2}s_{\text{int}er\sec t1} + \frac{1}{2}s_{\text{int}er\sec t2} \quad (16)$$

$$s_{\text{int}er\sec t1} = 2R^2 \cos^{-1}(\frac{d_1^2}{2Rd_1}) - \frac{1}{2}\sqrt{(-d_1+2R)d_1^2(d_1+2R)}$$

$$s_{\text{int}er\sec t2} = 2R^2 \cos^{-1}(\frac{d_2^2}{2Rd_2}) - \frac{1}{2}\sqrt{(-d_2+2R)d_2^2(d_2+2R)}$$

**3.f**

It is obviously that there is not hole.

**3.g**

It calculates like 1.d .

**3.h**

It calculates like 1 .a

**3.i**

In this state we have three intersections region so we have

(17)

$$s_h = s_\Delta - (s_1 + s_2 + s_3) + \frac{1}{2}s_{inter sec t1} + \frac{1}{2}s_{inter sec t2} + \frac{1}{2}s_{inter sec t3}$$

$$s_{inter sec t1} = 2R^2 \cos^{-1}(\frac{d_1^2}{2Rd_1}) - \frac{1}{2}\sqrt{(-d_1+2R)d_1^2(d_1+2R)}$$

$$s_{inter sec t2} = 2R^2 \cos^{-1}(\frac{d_2^2}{2Rd_2}) - \frac{1}{2}\sqrt{(-d_2+2R)d_2^2(d_2+2R)}$$

$$s_{inter sec t3} = 2R^2 \cos^{-1}(\frac{d_3^2}{2Rd_3}) - \frac{1}{2}\sqrt{(-d_3+2R)d_3^2(d_3+2R)}$$

In all of states if $s_h > 0$ then we conclude we have a hole.

## 4. Destination selection

In our proposed diagram (HSTT) After deciding the existence of a coverage hole and its size, a stationary node needs to decide the number of mobile nodes and the target locations of these mobile nodes to heal its holes .We want to aim maximum coverage so after calculate the area of hole we choice one of the circumcircle or incircle type. If the area is less than mobile sensor sensing region we use the circumcircle center for target location, if area of hole is larger than sensor sensing reign we use the incircle center for target location to aim maximum coverage.

The circumcenter of a triangle can be found as the intersection of the three perpendicular bisectors. (A perpendicular bisector is a line that forms a right angle with one of the triangle's sides and intersects that side at its midpoint.) This is because the circumcenter is equidistant from any pair of the triangle's points, and all points on the perpendicular bisectors are equidistant from those points of the triangle. It shows in Fig 6.

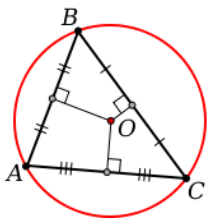

Fig.6 Construction of the circumcircle (red) and the circumcenter (red dot).

The incircle is the inscribed circle of a triangle $\Delta ABC$ i.e., the unique circle that is tangent to each of the triangle's three sides. The center $I$ of the incircle is called the incenter, and the radius $r$ of the circle is called the in radius. The incenter is the point of concurrence of the triangle's angle bisectors. In addition, the points $M_A$, $M_B$ and $M_C$ of intersection of the incircle with the sides of $\Delta ABC$ are the polygon vertices of the pedal triangle taking the incenter as the pedal point (c.f. tangential triangle). This triangle is called the contact triangle. It shows in fig 7.

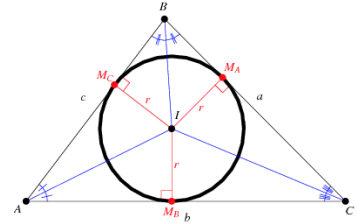

Fig.7 Construction of the incircle.

## 5. Conclusions

In this paper, we proposed a triangular oriented diagram (HSTT) that detect of hole and calculate its size. Also for maximum coverage we determine the target localization for mobile sensor for healing of coverage hole. All of them cause to along wireless network life time and optimization. In proposed diagram we use circumcircle and incircle to aim this goal. Our diagram compare with Voronoi diagram have some advantages such as

1. It is simple for construction.

2. It has lower than Voronoi diagram calculation for construction.

3. We get exact area of hole no estimation of it.

## 6. Future Work

We can emerge some adjacent triangular in our proposed diagram for use the low number of mobile sensor to healing coverage hole. Also we can use the hierarchical method according to size of hole to get maximum coverage by minimum mobile sensors. at the future we work to aim this goal.